\documentclass[%
 superscriptaddress,
 reprint,
 amsmath,amssymb,
aps,
pra,
]{revtex4-1}

\usepackage{graphicx}
\usepackage{subcaption}
\usepackage{caption}
 
\usepackage[textfont=normalfont,singlelinecheck=off,justification=raggedright ]{caption}

\usepackage{dcolumn}
\usepackage{bm}
\usepackage{amssymb}
\usepackage{amsmath}
\usepackage{txfonts}
\usepackage{mathdots}
\usepackage[classicReIm]{kpfonts}
\usepackage{xcolor}

\newcommand{\bra}[1] {\left\langle #1 \right|}
\newcommand{\ket}[1] {\left| #1 \right\rangle}
\def\openone{\leavevmode\hbox{\small1\kern-4.2pt\normalsize1}}

\newcommand{\nn}{\nonumber\\}
\newcommand{\bea}{\begin{eqnarray}}
\newcommand{\ea}{\end{eqnarray}}
\newcommand{\abs}[1]{\left|#1\right|}

\graphicspath{{./Images/}}
\usepackage[tableposition=top]{caption}
\usepackage{enumitem}

\begin{document}

\title{Realization of Surface Code Quantum Memory on Systems with Always-On Interactions}

\author{Sahar Daraeizadeh}
 \email{sahar2@pdx.edu}
 \affiliation{Department of Electrical and Computer Engineering, Portland State University}%
\author{Sarah Mostame}%
\affiliation{IBM T.J. Watson Research Center, Yorktown Heights, NY 10598, USA}
\author{Preethika Kumar Eslami}
\affiliation{Department of Electrical Engineering and Computer Science, Wichita State University, Wichita, KS 67260, USA}
\author{Xiaoyu Song}
\affiliation{Department of Electrical and Computer Engineering, Portland State University}%
\author{Marek Perkowski}
\affiliation{Department of Electrical and Computer Engineering, Portland State University}%

\date{\today}

\begin{abstract}
We realize Surface Code quantum memories for nearest-neighbor qubits with always-on Ising interactions. This is done by utilizing multi-qubit gates that mimic the functionality of several gates. The previously proposed Surface Code memories rely on error syndrome detection circuits based on CNOT gates. In a two-dimensional planar architecture, to realize a two-qubit CNOT gate in the presence of couplings to other neighboring qubits, the interaction of the target qubit with its three other neighbors must cancel out. Here we present a new error syndrome detection circuit utilizing multi-qubit parity gates. In addition to speed up in the error correction cycles, in our approach, the depth of the error syndrome detection circuit does not grow by increasing the number of qubits in the logical qubit layout. We analytically design the system parameters to realize new five-qubit gates suitable for error syndrome detection in nearest-neighbor two-dimensional array of qubits. The five-qubit gates are designed such that the middle qubit is the target qubit and all four coupled neighbors are the control qubits. In our scheme, only one control parameter of the target qubits must be adjusted to realize controlled-unitary operations. The gate operations are confirmed with a fidelity of $\mathrm{>}$99.9\% in a simulated system consists of nine nearest-neighbor qubits.
\end{abstract}

\pacs{Valid PACS appear here}

\maketitle

\section{Introduction}
\indent One of the most important areas of research in the field of quantum computing is to design and implement highly efficient and fault-tolerant scalable quantum architectures. The quantum systems are intrinsically error-prone since the states of qubits can change by environmentally-induced errors. Therefore, to realize a quantum memory, it is required to apply Quantum Error Correction (QEC) schemes \cite{m._2015} to preserve the states of the qubits during idle times. One of the most promising QEC schemes is Surface Code \cite{martinis_2015}.
\noindent The Surface Code architecture consists of Z and X stabilizers \cite{gottesman_1997, steane} and introduces ancillary qubits dedicated to these stabilizers. The code repetitively performs projective quantum non-demolition (QND) parity measurements on these ancillary qubits to measure the bit-flip and phase-flip errors of the data qubits \cite{fowler_mariantoni_martinis_cleland_2012}. The number of ancillary qubits in these measurements is approximately equal to the number of data qubits. Although it has been shown this approach results in storing information with a lower error rate, the Surface Code methodology has a high computational and resource overhead to realize the logical states and process the information. In this work, we propose a protocol to implement an efficient quantum memory based on Surface Code with applications in large scale 2-Dimensional (2D) nearest-neighbor (NN) quantum architectures with always-on interactions. This is possible due to our proposed five-qubit parity gates which can be applied in parallel on the entire array of qubits. 

\indent Parity gates can be used as an elementary gate in universal quantum computation \cite{ionicioiu_2007}. Kumar et. al. \cite{kumar_daraeizadeh_2015} designed a single-shot multi-qubit parity gate for quantum systems with Ising interactions. That can be utilized to generate efficient circuits for Mirror Inversion (MI) \cite{albanese_christandl_datta_ekert_2004,karbach_stolze_2005,raussendorf_2005} as a sequence of controlled-unitary operations between 2D nearest-neighbor qubits with tunable couplings. This method significantly increases the efficiency by lowering the computational overhead since the state transfer can be achieved in fewer computational steps without requiring ancillary qubits. Furthermore, there is not any dephasing from idle qubits since all the qubits are used in the MI operation as target or control qubits. However, the method is limited to 2D systems with tunable couplings. Although it is easier to perform multi-qubit gates in the systems with tunable couplings, there are some disadvantages such as increased circuit complexity and more noise introduction. We generalize the previous approach to design five-qubit controlled-unitary gates to realize parity gates in 2D nearest neighbor layouts with always-on interactions. 

\indent In our model, each five-qubit parity gate consists of one target qubit which is coupled to four adjacent control qubits where each control qubit can act as an active control qubit or a dummy qubit. In the case of five-qubit parity gate with four active control qubits, the gate operates so that the state of the target qubit is flipped when the XOR of all four adjacent qubits is one. In other words, if the four adjacent control qubits have even or odd parity, the state of the middle target qubit is preserved or inverted, respectively. However, in the case of five-qubit parity gate with two active control qubits, the state of the target qubit flips when two adjacent qubits (two active control qubits) have odd parity, while the other two adjacent qubits act as dummy qubits and have no effect on the gate operation. 

\begin{figure}[htp]
    \centering
    \includegraphics[width=1.65in, height=0.70in, keepaspectratio=false]{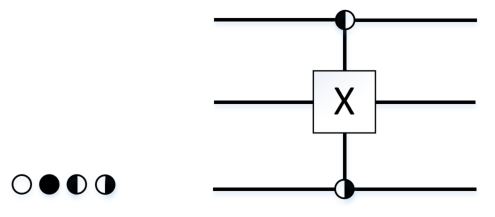}

\noindent                    (a)~~~~~~~~~~~~~~~~~~(b)    
    \caption{(a) Notations to represent the state of the control qubits in controlled unitary operations (b) A 3-qubits parity gate. Half colored circles can be used in pairs to represent the parity of the states of a pair of control qubits}
\label{Fig. 1}
\end{figure}

\indent Here we introduce a new symbol to represent the multi-qubit parity gates. As it is known, the symbol of a full-colored circle on a control qubit means when the logical state of control qubit is 1, the gate operation is performed on the target qubit. While the symbol of an empty circle means when the logical state of control qubit is 0, the gate operation is performed on the target qubit. We introduce the half-colored circles as shown in Fig.~\ref{Fig. 1}~(a) which means the logical state of the control qubit can be either 1 or 0. The half-colored circles are meaningful when applied in pairs to represent the opposite states of two control qubits resulting in a gate operation on target qubit. For example, in Fig.~\ref{Fig. 1}~(b) there is a left-half colored circle on top control qubit, while on the bottom control qubit there is a right-half colored circle. This means the two pairs of control qubits must be in opposite states for the target qubit to change its state (parity detection).

\section{Physical model and the simulation method}

\indent 
To simulate the dynamics of a quantum system, a time-dependent Schr\"{o}dinger equation  needs to be solved. Knowing the Hamiltonian of the system $H$ and the initial state $\ket{\Psi(t_0)}$, the time evolution of a quantum state is given by
$\ket{\Psi(t)} = U(t)\, \ket{\Psi(t_0)}\, $
with $U(t) =  e^{-iHt/\hbar}$ being the unitary transformation of the system. 
Throughout this paper, we consider $\hbar=1$. 

\indent Consider a $(m\times n)\,$ two-dimensional system of qubits with always-on nearest-neighbor Ising interactions. Such a system can be described by the following Hamiltonian\cite{raussendorf_2005, zhou_zhou_guo_feldman_2002,benjamin_bose_2003,fitzsimons_twamley_2006}, where the qubits are labeled with $j$ and $k$, for the rows and columns, respectively. 

\bea
\label{QubitHamiltonian} 
H &=& \sum^n_{k=1} \sum^m_{j=1}\left(\Delta_{j,k} \, \sigma_x ^{j,k}\, +\, \varepsilon_{j,k}\,\sigma_z ^{j,k} \right) + 
\nn
&+&  \sum^n_{k=1} \sum^{m-1}_{j=1} \xi_{j,j+1} ^k \,
\sigma_z ^{j,k} \, \sigma_z ^{j+1,k} 
+ 
\sum^{n-1}_{k=1} \sum^{m}_{j=1} \xi_j ^{k,k+1} \,
\sigma_z ^{j,k} \, \sigma_z ^{j,k+1}, \qquad
\ea
where $\sigma_{x}$ and $\sigma_{z}$ are Pauli operators,  $\Delta_{j,k}$ is the tunneling energy for the qubit located at the $j$-th row and $k$-th column, and $\varepsilon _{j,k}$ is the bias energy for the qubit. Here $\xi_{j,j+1} ^k $ is the coupling energy between two adjacent vertically coupled qubits in column $k$. Similarly, $\xi_j ^{k,k+1} $ is the coupling energy between two adjacent horizontally coupled qubits in row $j$. The Hamiltonian operator is a $2^{m\times n} \times2^{m\times n}$ matrix, which scales exponentially with the number of qubits in the system. It is challenging to solve such a large matrix analytically in order to derive the system parameters. However, using a pulses bias scheme \cite{kumar_skinner_Behrman_steck_Han_2005, kumar_skinner_2007} and reduced Hamiltonian technique \cite{kunmar_skinner_daraeizadeh_2011}, we can solve the system parameters to realize a desired multi-qubit parity gate. 

\indent  We consider a system of nine qubits as depicted in the black square in Fig. 2, where each qubit is interacting with 4 neighbors. We design a  controlled-unitary gate where qubits A, B, C, and D are control qubits and T is the target qubit. In the architecture shown in Fig. 2, we consider four coupling strengths ${\xi }_{\rm A}, \ {\xi }_{\rm B}, \ {\xi }_{\rm C}, \ {\rm and} \ {\xi }_{\rm D}$ respectively between the target qubit T and the control qubits A, B, C, and D. By design, the coupling strengths between pairs of qubits are alternating in a row or column of the two-dimensional array of qubits. Therefore, if any qubit in the array be selected as the Target qubit, it is interacting with four neighbors with four distinct coupling strength.

\indent The evolution of a nine-qubit system, qubits A, B, C, D, E, F, G, H and T in Fig. 2, is described by a $512\times512$ Hamiltonian matrix. Qubits E, F, G, and H have direct interactions with control qubits but do not have any direct interaction with the target qubit. In order to study their impact on the dynamics of the system, we applied the same bias parameters on E, F, G, and H qubits as for control qubits A, B, C, and D.  We observed that they do not affect the 5-qubit gate operation nor the gate operation affects the state of these qubits. 
Therefore, to find the parameters of a five-qubit gate operation on A, B, C, D and T, we analyze a $32\times32$ Hamiltonian matrix.
Using the reduced Hamiltonian scheme \cite{kunmar_skinner_daraeizadeh_2011,kumar_2012}, we break this Hamiltonian matrix to sixteen $2\times2$ Hamiltonian matrices. Each $2\times2$ Hamiltonian describes evolution of the target qubit T in a subspace depending on the states of the control qubits. 
Then for each of these $2\times 2$ Hamiltonians, we generate a unitary matrix by integrating the Schr\"{o}dinger equation, and then equating the generated unitary matrix to a desirable controlled unitary gate operation for that subspace. Next, we describe this in details. 

\begin{figure}[htp]
    \centering
    \includegraphics[width=2.2in, height=2.2in, keepaspectratio=false]{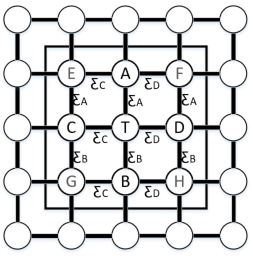}
    \caption{In non-tunable coupling systems, each qubit is interacting with 4 neighbor qubits. In this figure, we are interested to perform a Parity gate on the target qubit T where it is directly coupled to qubits A, B, C, and D.}
    \label{Fig. 2}
\end{figure}

\indent
The evolution of the target qubit T being directly coupled to the control qubits A, B, C, and D can be described by the reduced Hamiltonian:
\bea
\label{ReducedHamiltonian}
H_{\rm red} &=&
\Delta_{\rm T}\, \sigma_x^{\rm T}\,  \, + \, \left\{\varepsilon_{\rm T}\, +\,
\xi_{\rm A}\bra{\Phi}\sigma_z^{\rm A}\ket{\Phi} \, +\,
\xi_{\rm B}\bra{\Phi}\sigma_z^{\rm B}\ket{\Phi} \right.
\nn
&+& \left.  
\xi_{\rm C}\bra{\Phi}\sigma_z^{\rm C}\ket{\Phi} \, +\,
\xi_{\rm D}\bra{\Phi}\sigma_z^{\rm D}\ket{\Phi}
\right\} \,  \sigma_z^{\rm T}
\ea
where the label "red" stands for the reduced evolution subspace and $\ket\Phi$  represents the initial state of four control qubits A, B, C, and D, each is initialized to $\ket 0$ or $\ket 1$.
The parameters of the Hamiltonian are the same as Eq.~(\ref{QubitHamiltonian}) but for simplicity, we have dropped some labels for the 5-qubits system shown in Fig.~\ref{Fig. 2}: $\Delta_{\rm T}$, $\varepsilon_{\rm T}$ are the tunneling energy and the bias energy for the target qubit, respectively, and $\xi_i$ is the coupling energy between control qubit ``$i$'' and the target qubit, where $i$ = A, B, C, D. Depending on the initial state of $\ket\Phi$, the expectation value of $\sigma_z^i$ can be +1 or -1.

\indent Note that here we have ignored the effect of the next-nearest-neighbor couplings. 
The qubits E, F, G, and H do not contribute to the evolution of the target qubit T as they don't have any direct coupling with T. In our simulations, the biases on non-interacting qubits E, F, G, and H are set such that their states are preserved. 

\noindent 
Given the above reduced Hamiltonian, the unitary transformation on target qubit in terms of the system parameters can be derived as 
{\small 
\bea
\label{TUnitary} 
&& U (t)= e^{i\theta } 
\left[\begin{array}{cc}
\cos(\omega t) - \frac{2\pi i E}{\omega}\sin(\omega t) & 
\frac{2\pi (- i \Delta_{\rm T})}{\omega}\, \sin(\omega t) \\ 
& 
\\
\frac{2\pi (- i \Delta_{\rm T})}{\omega}\, \sin(\omega t) 
& 
\cos(\omega t) + \frac{2\pi i E}{\omega}\sin(\omega t)
\end{array} 
\right] \, ,
\nn
\ea}
\hspace{-0.3cm}
with $E$ being the effective bias and $\omega = 2\pi \sqrt{\Delta_{\rm T}^2 + E^2}$ \, being the angular momentum of the gate operation. 
Here $\theta $ is a global phase factor.

\indent  Designing multiple-controlled unitary gates in a system with always-on interactions requires careful attention to the connectivity/couplings between qubits. To design a controlled-unitary gate where the target qubit is interacting with a set of neighbors but only a subset of neighbors have a control role; one needs to cancel out the effect of those neighbors who do not have a control role. For instance, in Fig. 2 consider designing a CNOT gate between qubits A and T, where A is the control qubit and T is the target qubit. Here, the qubits B, C, and D have direct interaction with qubit T but do not have a control role. Therefore, we need to design a five-qubit controlled-unitary gate with one target qubit T, one active control qubit A, and three dummy qubits B, C, and D. Note that the states of the dummy qubits should not effect the CNOT gate operation between A and T. Since there are 8 logical states ($000, 001, ..., 111$) associated with the dummy qubits, to achieve the desired CNOT gate, one needs to realize a sequence of 8 five-qubit controlled-unitary operations, each taking the duration of $\tau$. Where each five-qubit controlled-unitary operation configures T as the target qubit, A as the control qubit with logical state 1, and B, C, and D as the control qubits with one of the 8 logical states \cite{kunmar_skinner_daraeizadeh_2011}. 

\indent In most of the error correction codes such as Repetition Code \cite{kelly_2015}, the bit-flip error syndrome detection circuit uses a sequence of two CNOT gates applied on two data qubits as control qubits and one measurement qubit as the target qubit. In a 2-dimensional system with always-on interaction, this results in a decomposition to a sequence of sixteen controlled-unitary operations (16$\tau$). Utilizing the five-qubit parity gates with two active control qubits, we realize the same functionality while reducing the circuit depth to a sequence of only three controlled-unitary operations (3$\tau$).
\section{Five-qubit parity gates with two active control qubits}
\label{ParityGateTwoControlQ}

\begin{figure}
    \includegraphics[width=2.92in, height=1.46in, keepaspectratio=false]{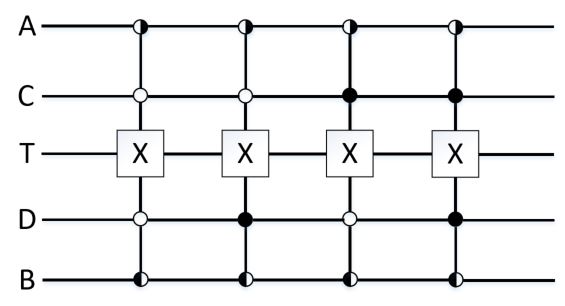}

 (a)
 
    \includegraphics[width=2.92in, height=1.46in, keepaspectratio=false]{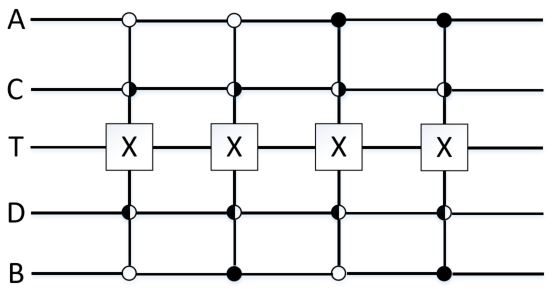}

 (b)
 \caption{\label{Fig. 3}The circuits to realize parity gates with only two active vertical~(a) or horizontal~(b) control qubits. }
 \end{figure}

\indent  
In this section we design a five-qubit parity gate in a 2D array of qubits where only two of the four control qubits have an active effect while the effects of two other control qubits are canceled.
As discussed above and shown in Fig.~\ref{Fig. 2}, qubit T is the target qubit.
Our goal here is to apply a parity gate to detect the parity of qubits A and B which are vertically coupled to the target qubit T. Therefore, we perform an $X$ unitary operation on the target qubit $U_{\rm T}  = X$ in the subspaces
$Q_{\rm A}Q_{\rm B} = \ket{10}$ and $Q_{\rm A}Q_{\rm B} = \ket{01}$, irrespective of the states of the qubits C, and D. 
In the subspaces where $Q_{\rm A}Q_{\rm B} = \ket{00}$ 
or
$Q_{\rm A}Q_{\rm B} = \ket{11}$  we will perform an Identity unitary operation on the target qubit 
$U_{\rm T}  = I$. 
This is done by applying a sequence of four controlled-unitary gates as shown in Fig.~\ref{Fig. 3}~(a). 
Similarly, the circuit shown in Fig.~\ref{Fig. 3}~(b) realizes a parity detector gate where the qubits C and D are the two actively effective control qubits.

{\begin{figure}
\centering
    \includegraphics[width=2.63in]{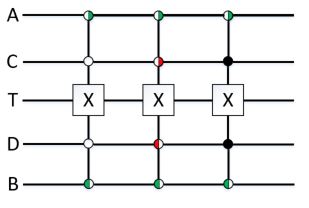}

(a)

    \includegraphics[height=1.64in]{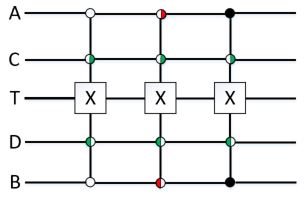}
    
 (b)

\caption{
(a)~Two active vertical control qubits. The red half-colored circles represent the different states of qubits C, and D.
(b)~Two active horizontal control qubits. The red half-colored circles represent the different states of qubits A, and B. The same principle applies to green half-colored circles.
}
\label{Fig. 4}
\end{figure}}

\noindent In Fig.~\ref{Fig. 3}~(a) and (b), the two gates located in the middle can be combined into one gate which operates on the target qubit if the qubits A vs B and C vs D are in different states. 
We can represent this gate with four half colored circles on the control qubits, where (A, B) and (C, D) are considered the pairs with opposite half-colored circles as shown in Fig.~\ref{Fig. 4}, where the pairs of half-colored circles are color-coded.

\begin{table}
\caption{Effective bias under each subspace when qubit T is coupled to four neighbor qubits A, B, C, D.}
\centering
\begin{tabular}{|p{0.2in}|p{0.6in}|p{0.6in}|p{1.6in}|} \hline 
 & AB & CD & Effective Bias \\ \hline 
0 & $\left.|00\right\rangle $ & $\left.|00\right\rangle $ & ${E=\varepsilon }_T+{\xi }_{\rm A}+{\xi }_{\rm B}+\ {\xi }_{\rm C}+{\xi }_{\rm D}$ \\ \hline 
1 & $\left.|00\right\rangle $ & $\left.|01\right\rangle $ & ${E=\varepsilon }_T+{\xi }_{\rm A}+{\xi }_{\rm B}+\ {\xi }_{\rm C}-{\xi }_{\rm D}$ \\ \hline 
2 & $\left.|00\right\rangle $ & $\left.|10\right\rangle $ & ${E=\varepsilon }_T+{\xi }_{\rm A}+{\xi }_{\rm B}-\ {\xi }_{\rm C}+{\xi }_{\rm D}$ \\ \hline 
3 & $\left.|00\right\rangle $ & $\left.|11\right\rangle $ & ${E=\varepsilon }_T+{\xi }_{\rm A}+{\xi }_{\rm B}-\ {\xi }_{\rm C}-{\xi }_{\rm D}$ \\ \hline 
4 & $\left.|01\right\rangle $ & $\left.|00\right\rangle $ & ${E=\varepsilon }_T+{\xi }_{\rm A}-{\xi }_{\rm B}+\ {\xi }_{\rm C}+{\xi }_{\rm D}$ \\ \hline 
5 & $\left.|01\right\rangle $ & $\left.|01\right\rangle $ & ${E=\varepsilon }_T+{\xi }_{\rm A}-{\xi }_{\rm B}+\ {\xi }_{\rm C}-{\xi }_{\rm D}$ \\ \hline 
6 & $\left.|01\right\rangle $ & $\left.|10\right\rangle $ & ${E=\varepsilon }_T+{\xi }_{\rm A}-{\xi }_{\rm B}-\ {\xi }_{\rm C}+{\xi }_{\rm D}$ \\ \hline 
7 & $\left.|01\right\rangle $ & $\left.|11\right\rangle $ & ${E=\varepsilon }_T+{\xi }_{\rm A}-{\xi }_{\rm B}+\ {\xi }_{\rm C}-{\xi }_{\rm D}$ \\ \hline 
8 & $\left.|10\right\rangle $ & $\left.|00\right\rangle $ & ${E=\varepsilon }_T-{\xi }_{\rm A}+{\xi }_{\rm B}+\ {\xi }_{\rm C}+{\xi }_{\rm D}$ \\ \hline 
9 & $\left.|10\right\rangle $ & $\left.|01\right\rangle $ & ${E=\varepsilon }_T-{\xi }_{\rm A}+{\xi }_{\rm B}+\ {\xi }_{\rm C}-{\xi }_{\rm D}$ \\ \hline 
10 & $\left.|10\right\rangle $ & $\left.|10\right\rangle $ & ${E=\varepsilon }_T-{\xi }_{\rm A}+{\xi }_{\rm B}-\ {\xi }_{\rm C}+{\xi }_{\rm D}$ \\ \hline 
11 & $\left.|10\right\rangle $ & $\left.|11\right\rangle $ & ${E=\varepsilon }_T-{\xi }_{\rm A}+{\xi }_{\rm B}-\ {\xi }_{\rm C}-{\xi }_{\rm D}$ \\ \hline 
12 & $\left.|11\right\rangle $ & $\left.|00\right\rangle $ & ${E=\varepsilon }_T-{\xi }_{\rm A}-{\xi }_{\rm B}+\ {\xi }_{\rm C}+{\xi }_{\rm D}$ \\ \hline 
13 & $\left.|11\right\rangle $ & $\left.|01\right\rangle $ & ${E=\varepsilon }_T-{\xi }_{\rm A}-{\xi }_{\rm B}+\ {\xi }_{\rm C}-{\xi }_{\rm D}$ \\ \hline 
14 & $\left.|11\right\rangle $ & $\left.|10\right\rangle $ & ${E=\varepsilon }_T-{\xi }_{\rm A}-{\xi }_{\rm B}-\ {\xi }_{\rm C}+{\xi }_{\rm D}$ \\ \hline 
15 & $\left.|11\right\rangle $ & $\left.|11\right\rangle $ & ${E=\varepsilon }_T-{\xi }_{\rm A}-{\xi }_{\rm B}-\ {\xi }_{\rm C}-{\xi }_{\rm D}$ \\ \hline 
\end{tabular}
\label{Table Effective Bias}
\end{table}

\indent Note we consider only the qubits that have direct coupling with the target T. In Table \ref{Table Effective Bias} we list the effective bias in all 16 possible subspaces of 
$Q_{\rm A} Q_{\rm B}=\ket{10}$, 
$Q_{\rm A}Q_{\rm B} =\ket{01}$, 
$Q_{\rm A}Q_{\rm B} =\ket{00}$, 
and 
$Q_{\rm A}Q_{\rm B} =\ket{11}$, 
where qubits C, and D, have arbitrary values. 

\noindent In the subspace ${Q}_{\rm A}{Q}_{\rm B}=\ket{10}$, where qubits C, and D have arbitrary values, the effective bias is 
\mbox{$
E={\varepsilon }_{\rm {T}}\, -\, {\xi }_{\rm {A}}+{\xi }_{\rm{B}}\pm{\xi }_{\rm {C}}\pm {\xi }_{\rm{D}} 
$.}
To realize an $X$ operation, we need to cancel the diagonal terms in Eq.~(\ref{TUnitary}), and force $\sin (\omega t)=1$ and
$\omega = 2\pi \Delta_{\rm T}$ which results in 
$-2\pi i\,\Delta_{\rm T}\, \sin(\omega t)/\omega = -i$, 
where $-i$  contributes as a phase factor of $3\pi /2$ on the target qubit. 
This extra phase on the target qubit can be tracked in the course of computation. The following conditions must be satisfied:
\bea
\label{cos-sin=0} 
&& \cos (\omega t)\, -\, \frac{2\pi i E}{\omega }\, \sin (\omega t)  = 
\, \cos (\omega t)+ \frac{2\pi i E}{\omega}\, \sin (\omega t) = 0
\nn
&& \hspace{2.5cm} \Longrightarrow  E=0 , \quad  \cos (\omega t)=0 
\ea
Considering $t=\tau$ for the $X$ operation time, we need to satisfy condition:
\mbox{$\omega \tau = (4n+1) \pi/2 $},
where n is an integer. 
 For rf-SQUID (Superconducting Quantum Interference Devices) qubit systems, one set of parameters which satisfies the conditions above would be 
 \mbox{$\Delta_{\rm T} = 25$~MHz}, $n=0$, and $\tau = 10$~ns, while bias pulse magnitude can range up to 10 GHz \cite{kumar_skinner_Behrman_steck_Han_2005, kunmar_skinner_daraeizadeh_2011, kumar_daraeizadeh_2015}. Canceling out the effective bias $(E=0)$ we would also need the condition:
\bea
\label{ConTbias} 
{\varepsilon }_{\rm {T}}={\xi }_{\rm {A}}-{\xi }_{\rm{B}} \pm {\xi }_{\rm{C}}\pm {\xi }_{\rm{D}} \, .
\ea
As shown in Table \ref{Table Effective Bias}, the effective bias in Eq.~(\ref{ConTbias}) -- subspace 
$Q_{\rm A} Q_{\rm B}=\ket{10}$ -- expands to four subspaces depending on the state of ${Q}_{\rm C}{Q}_{\rm D}$:
\bea
\label{ConCD} 
{Q}_{\rm C}{Q}_{\rm D} &=&\left.\mathrm{|00}\right\rangle, \ {\varepsilon }_{\rm T}={\xi }_{\mathrm{A}}-{\xi }_{\mathrm{B}}-\ {\xi }_{\mathrm{C}}-{\xi }_{\mathrm{D}} 
\nn
{Q}_{\rm C}{Q}_{\rm D} &=&\left.\mathrm{|01}\right\rangle, \ {\varepsilon }_{\rm T}={\xi }_{\rm{A}}-{\xi }_{\rm{B}}-{\xi }_{\rm{C}}+{\xi }_{\mathrm{D}}                  
\nn
{Q}_{\rm C}{Q}_{\rm D} &=&\left.\mathrm{|10}\right\rangle, \ {\varepsilon }_{\rm T}={\xi }_{\mathrm{A}}-{\xi }_{\mathrm{B}}+\ {\xi }_{\mathrm{C}}-{\xi }_{\mathrm{D}}     
\nn 
{Q}_{\rm C}{Q}_{\rm D} &=&\left.\mathrm{|11}\right\rangle, \ {\varepsilon }_{\rm T}={\xi }_{\mathrm{A}}-{\xi }_{\mathrm{B}}+\ {\xi }_{\mathrm{C}}+{\xi }_{\mathrm{D}} 
\, .
\ea
\noindent By choosing ${\xi }_{\rm{A}}={\xi }_{\rm{B}}$ and ${\xi }_{\rm{C}}={\xi }_{\mathrm{D}}$, for 
\mbox{$Q_{\rm C} Q_{\rm D} = \ket{01}$}, $\ket{10}$  we get ${\varepsilon }_{\rm T}=0$, while for 
\mbox{$Q_{\rm C} Q_{\rm D} =\ket{00}$,} $\ket{11}$ 
we have 
\mbox{${\varepsilon }_{\rm T}=-{\xi }_{\rm{C}}-{\xi }_{\rm{D}}=-2{\xi }_{\rm{D}}$,} and ${\varepsilon }_{\rm T}={\xi }_{\rm{C}}+{\xi }_{\rm{D}}=2{\xi }_{\rm{D}}$. 
\vspace{0.2cm}

\indent
A similar calculation can be done for subspace \mbox{${Q}_{\rm A}{Q}_{\rm B}=\left.\mathrm{|01}\right\rangle $} with the same results for the  bias on the target qubit  ($\varepsilon_{\rm T}$).
Therefore to realize an $X$ operation on the target qubit, we keep biases on all control qubits at some arbitrary value such that it would not cancel the effect of couplings \cite{kunmar_skinner_daraeizadeh_2011} \mbox{${\varepsilon }_{\rm{A}}={\varepsilon }_{\rm{B}}=\ {\varepsilon }_{\rm{C}}={\varepsilon }_{\rm{D}}=2$} GHz, and apply a sequence of  bias pulse steps on the target qubit as following, with each of them taking $\tau=10$~ns:

\bea
{\varepsilon }_{{\rm T}_{1}}\ = -{\xi }_{\rm{C}}-{\xi }_{\rm{D}}, \quad
{\varepsilon }_{{\rm T}_{2}}\ = 0,
\quad
{\varepsilon }_{{\rm T}_{3}}\ ={\xi }_{\rm{C}}+{\xi }_{\rm{D}}\
\label{VerticalPulse1} 
\ea
where ${\varepsilon }_{{\rm T}_i}$ represents the ${i}$-th bias magnitude on target qubit T. The order of applying these three pulse steps does not matter, since at the end after 30 ns, the desired gate operation has been realized. 
Table \ref{Table Effective Bias Pulses} summarizes all possible effective biases in each subspace under the three pulse steps given by Eq.~(\ref{VerticalPulse1}). 
This table is derived by substituting the bias magnitude of the target qubit under each pulse step (${\varepsilon }_{{\rm T}_{1}}\mathrm{,\ }{\varepsilon }_{{\rm T}_{2}}\mathrm{,\ }{\varepsilon }_{{\rm T}_{3}}$) in the effective bias $E$ given in Table \ref{Table Effective Bias}.

\begin{table*}
\centering
\caption{Effective bias in each subspace under each pulse sequence}
\begin{tabular*}{\textwidth}{|p{1in}|p{1in}|p{1.6in}|p{1.6in}|p{1.6in}|} \hline 
AB & CD & ${\varepsilon }_{T_1}=-{\xi }_{\rm C}-{\xi }_{\rm D}$ & ${\varepsilon }_{T_2}=0$ & ${\varepsilon }_{T_3}={\xi }_{\rm C}+{\xi }_{\rm D}$ \\ \hline 
$\left.|00\right\rangle $ & $\left.|00\right\rangle $ & $E=+{\xi }_{\rm A}+{\xi }_{\rm B}$ & $E={\xi }_{\rm A}+{\xi }_{\rm B}+\ {\xi }_{\rm C}+{\xi }_{\rm D}$ & $E={\xi }_{\rm A}+{\xi }_{\rm B}+\ {2{\xi }_{\rm C}}+{2{\xi }_{\rm D}}$ \\ \hline 
$\left.|00\right\rangle $ & $\left.|01\right\rangle $ & $E={\xi }_{\rm A}+{\xi }_{\rm B}-2{\xi }_{\rm D}$ & $E={\xi }_{\rm A}+{\xi }_{\rm B}+\ {\xi }_{\rm C}-{\xi }_{\rm D}$ & $E={\xi }_{\rm A}+{\xi }_{\rm B}+\ {2{\xi }_{\rm C}}$ \\ \hline 
$\left.|00\right\rangle $ & $\left.|10\right\rangle $ & $E={\xi }_{\rm A}+{\xi }_{\rm B}-\ {2{\xi }_{\rm C}}$ & $E={\xi }_{\rm A}+{\xi }_{\rm B}-\ {\xi }_{\rm C}+{\xi }_{\rm D}$ & $E={\xi }_{\rm A}+{\xi }_{\rm B}+\ {2{\xi }_{\rm D}}$ \\ \hline 
$\left.|00\right\rangle $ & $\left.|11\right\rangle $ & $E={\xi }_{\rm A}+{\xi }_{\rm B}-\ {2{\xi }_{\rm C}}-{2{\xi }_{\rm D}}$ & $E={\xi }_{\rm A}+{\xi }_{\rm B}-\ {\xi }_{\rm C}-{\xi }_{\rm D}$ & $E=+{\xi }_{\rm A}+{\xi }_{\rm B}$ \\ \hline 
$\left.|01\right\rangle $ & $\left.|00\right\rangle $ & $E=+{\xi }_{\rm A}-{\xi }_{\rm B}$ & $E={\xi }_{\rm A}-{\xi }_{\rm B}+\ {\xi }_{\rm C}+{\xi }_{\rm D}$ & $E={\xi }_{\rm A}-{\xi }_{\rm B}+\ 2{\xi }_{\rm C}+ 2{\xi }_{\rm D}$ \\ \hline 
$\left.|01\right\rangle $ & $\left.|01\right\rangle $ & $E={\xi }_{\rm A}-{\xi }_{\rm B}-2{\xi }_{\rm D}$ & $E={\xi }_{\rm A}-{\xi }_{\rm B}+\ {\xi }_{\rm C}-{\xi }_{\rm D}$ & $E={\xi }_{\rm A}-{\xi }_{\rm B}+\ 2{\xi }_{\rm C}$ \\ \hline 
$\left.|01\right\rangle $ & $\left.|10\right\rangle $ & $E={\xi }_{\rm A}-{\xi }_{\rm B}-\ 2{\xi }_{\rm C}$ & $E={\xi }_{\rm A}-{\xi }_{\rm B}-\ {\xi }_{\rm C}+{\xi }_{\rm D}$ & $E={\xi }_{\rm A}-{\xi }_{\rm B}+\ 2{\xi }_{\rm D}$ \\ \hline 
$\left.|01\right\rangle $ & $\left.|11\right\rangle $ & $E={\xi }_{\rm A}-{\xi }_{\rm B}-\ 2{\xi }_{\rm C}-2{\xi }_{\rm D}$ & $E={\xi }_{\rm A}-{\xi }_{\rm B}-\ {\xi }_{\rm C}-{\xi }_{\rm D}$ & $E=+{\xi }_{\rm A}-{\xi }_{\rm B}$ \\ \hline 
$\left.|10\right\rangle $ & $\left.|00\right\rangle $ & $E=-{\xi }_{\rm A}+{\xi }_{\rm B}$ & $E=-{\xi }_{\rm A}+{\xi }_{\rm B}+\ {\xi }_{\rm C}+{\xi }_{\rm D}$ & $E=-{\xi }_{\rm A}+{\xi }_{\rm B}+\ 2{\xi }_{\rm C}+2{\xi }_{\rm D}$ \\ \hline 
$\left.|10\right\rangle $ & $\left.|01\right\rangle $ & $E=-{\xi }_{\rm A}+{\xi }_{\rm B}-2{\xi }_{\rm D}$ & $E=-{\xi }_{\rm A}+{\xi }_{\rm B}+\ {\xi }_{\rm C}-{\xi }_{\rm D}$ & $E=-{\xi }_{\rm A}+{\xi }_{\rm B}+\ 2{\xi }_{\rm C}$ \\ \hline 
$\left.|10\right\rangle $ & $\left.|10\right\rangle $ & $E=-{\xi }_{\rm A}+{\xi }_{\rm B}-\ 2{\xi }_{\rm C}$ & $E=-{\xi }_{\rm A}+{\xi }_{\rm B}-\ {\xi }_{\rm C}+{\xi }_{\rm D}$ & $E=-{\xi }_{\rm A}+{\xi }_{\rm B}+\ 2{\xi }_{\rm D}$ \\ \hline 
$\left.|10\right\rangle $ & $\left.|11\right\rangle $ & $E=-{\xi }_{\rm A}+{\xi }_{\rm B}-\ 2{\xi }_{\rm C}-2{\xi }_{\rm D}$ & $E=-{\xi }_{\rm A}+{\xi }_{\rm B}-\ {\xi }_{\rm C}-{\xi }_{\rm D}$ & $E=-{\xi }_{\rm A}+{\xi }_{\rm B}$ \\ \hline 
$\left.|11\right\rangle $ & $\left.|00\right\rangle $ & $E=-{\xi }_{\rm A}-{\xi }_{\rm B}$ & $E=-{\xi }_{\rm A}-{\xi }_{\rm B}+\ {\xi }_{\rm C}+{\xi }_{\rm D}$ & $E=-{\xi }_{\rm A}-{\xi }_{\rm B}+\ 2{\xi }_{\rm C}+2{\xi }_{\rm D}$ \\ \hline 
$\left.|11\right\rangle $ & $\left.|01\right\rangle $ & $E=-{\xi }_{\rm A}-{\xi }_{\rm B}-2{\xi }_{\rm D}$ & $E=-{\xi }_{\rm A}-{\xi }_{\rm B}+\ {\xi }_{\rm C}-{\xi }_{\rm D}$ & $E=-{\xi }_{\rm A}-{\xi }_{\rm B}+\ 2{\xi }_{\rm C}$ \\ \hline 
$\left.|11\right\rangle $ & $\left.|10\right\rangle $ & $E=-{\xi }_{\rm A}-{\xi }_{\rm B}-\ 2{\xi }_{\rm C}$ & $E=-{\xi }_{\rm A}-{\xi }_{\rm B}-\ {\xi }_{\rm C}+{\xi }_{\rm D}$ & $E=-{\xi }_{\rm A}-{\xi }_{\rm B}+\ 2{\xi }_{\rm D}$ \\ \hline 
$\left.|11\right\rangle $ & $\left.|11\right\rangle $ & $E=-{\xi }_{\rm A}-{\xi }_{\rm B}-\ 2{\xi }_{\rm C}-2{\xi }_{\rm D}$ & $E=-{\xi }_{\rm A}-{\xi }_{\rm B}-\ {\xi }_{\rm C}-{\xi }_{\rm D}$ & $E=-{\xi }_{\rm A}-{\xi }_{\rm B}$ \\ \hline 
\end{tabular*}
\label{Table Effective Bias Pulses}
\end{table*}

\indent Then to perform an $X$ operation in subspaces where \mbox{${Q}_{\rm A}{Q}_{\rm B}=\left.\mathrm{|10}\right\rangle $} and \mbox{${Q}_{\rm A}{Q}_{\rm B}=\left.\mathrm{|01}\right\rangle$}, we set the coupling values such that the effective bias is canceled out under one of the three pulse steps (${\varepsilon }_{{\rm T}_{1}}\mathrm{,\ }{\varepsilon }_{{\rm T}_{2}}\mathrm{,\ }{\varepsilon }_{{\rm T}_{3}}$) while an Identity operation is realized elsewhere (see Table \ref{Table Effective Bias Pulses}). 
For all other subspaces where 
\mbox{${Q}_{\rm A}{Q}_{\rm B}=\left.\mathrm{|00}\right\rangle $}
and 
\mbox{${Q}_{\rm A}{Q}_{\rm B}=\left.\mathrm{|11}\right\rangle $}, we want to achieve Identity operation under all three pulse steps (${\varepsilon }_{{\rm T}_{1}}\mathrm{,\ }{\varepsilon }_{{\rm T}_{2}}\mathrm{,\ }{\varepsilon }_{{\rm T}_{3}}$). By choosing ${\xi }_{\rm{A}}={\xi }_{\rm{B}}$ and ${\xi }_{\rm{C}}={\xi }_{\rm{D}}$, most of the equations in Table \ref{Table Effective Bias Pulses} simplify or cancel out and only 7 effective bias equations remain which are listed below 
\bea 
\label{Eq. 17} 
E&=&2{\xi }_{\rm{B}}, \quad        
E=2{\xi }_{\rm{D}}, \quad
E=4{\xi }_{\rm{D}}, \quad
E=2{\xi }_{\rm{B}}+2{\xi }_{\rm{D}} 
\nn 
E&=&2{\xi }_{\rm{B}}-2{\xi }_{\rm{D}}, \quad
E=2{\xi }_{\rm{B}}+4{\xi }_{\rm{D}}, \quad 
E=2{\xi }_{\rm{B}}-4{\xi }_{\rm{D}} 
\ea
Under the above effective biases, we like to achieve an Identity gate operation. Therefore, we should choose the coupling values such that the off-diagonal terms in Eq.~\ref{TUnitary} are zero and diagonal terms are 1.
This results in 
\bea 
\label{ConCos=1} 
cos \left(\omegaup t\right)\ =1 \quad \Rightarrow \quad \omega=\frac{2\pi n}{\tau} 
\ea 
where $n$ is an integer. For $\xi \gg \Delta_{\rm T}$, we can ignore $\Delta_{\rm T}^{2}$ in 
\mbox{$\omega =2\pi \sqrt{\Delta_{\rm T}^{2}+E^2}$}, which results in 
\mbox{$\omega =2\pi E=2\pi n/\tau$}. Therefore, we choose the effective biases in the equations above as multiples of some integers $\tau$ such that
\bea
\label{Eq. 25} 
&& 
{2\xi }_{\rm B} = 2\frac{v}{\tau}, \quad
{2\xi }_{\rm D}=2\frac{w}{\tau}, \quad
4\xi_{\rm D}=4\frac{w}{\tau} \quad \Rightarrow
\nn
&&
2{\xi }_{\rm{B}}\pm 2{\xi }_{\rm{D}}=2\frac{v\pm w}{\tau}, 
\quad
{2\xi }_{\rm B}\pm 4{\xi }_{\rm{D}}=2\frac{v\pm 2w}{\tau} 
\ea 

\noindent where $v$ and $w$ are integers. One set of values for a system with tunneling energy $\Delta_{\rm T}=25$~MHz and $\tau = 10$~ns are the coupling values 
\mbox{${\xi }_{\rm A}={\xi }_{\rm B}=0.6$~GHz} and 
\mbox{${\xi }_{\rm C}={\xi }_{\rm D}=0.4$~GHz}. The above set of parameters realizes a parity gate that detects the parity of qubits A vs B (vertical) while ignoring the states of D and C (horizontal). 
 Similar calculations can be done to design a parity gate that detects the parity of qubits D vs C (horizontal) while ignoring the states of A and B (vertical). 
 Here, we would like to perform an $X$ unitary operation on the target qubit $\left ({U}_{\rm T} = X \right)$ in subspaces ${Q}_{\rm C}{Q}_{\rm D}=\left.\rm{|10}\right\rangle $ and ${Q}_{\rm C}{Q}_{\rm D}=\left.\rm{|01}\right\rangle $, no matter what the states of qubits A and B are. In all other subspaces where ${Q}_{\rm C}{Q}_{\rm D }=\left.\rm{|00}\right\rangle $ or ${Q}_{\rm C}{Q}_{\rm D }=\left.\rm{|11}\right\rangle $, we want to perform an Identity operation on qubit T $\left ({U}_{\rm T} = I \right)$. Using the set of parameters as discussed previously, we can apply a bias pulse on the target qubit with the following three magnitudes, each taking $\tau=10$~ns
\bea 
\label{Eq. 29} 
{\varepsilon }_{{\rm T}_{1}}=-{\xi }_{\rm A}-{\xi }_{\rm B},
\quad
{\varepsilon }_{{\rm T}_{2}}=0,
\quad
{\varepsilon }_{{\rm T}_{3}}={\xi }_{\rm A}+{\xi }_{\rm B} 
\ea 

\section{Five-qubit parity gate with four active control qubits} \label{ParityGateFourControlQ}

\indent Now consider the case where the target qubit detects the even or odd parity of the four control qubits. For the target qubit to flip when the parity of four control qubits is odd, we need to treat all four control qubits equally, therefore, we set the coupling values connected to the target qubit equal ${\xi }_{\rm A}={\xi }_{\rm B}= {\xi }_{\rm C}= {\xi }_{\rm D} = \xi $. To realize an Identity operation on subspaces when the parity of four control qubits is even (subspaces in rows 0, 3, 5, 6, 9, 10, 12, and 15 from Table \ref{Table Effective Bias}), the effective bias on the target qubit must be chosen such that the angular frequency of the target qubit equals an integer multiple of $ 2\pi $ over the Identity operation duration, say \mbox{$\omega =2\pi E=2\pi n/\tau$}, with $n$ being an integer. This results in 
\bea
\label{Eq. 32} 
E =\varepsilon_{\rm T} = \frac{v}{\tau}, \quad
E =\varepsilon_{\rm T} \pm \ 4\xi  = \frac{w}{\tau}          
\ea

\noindent where $v$ and $w$ are integers. Using the same parameters derived in the previous section for the initial bias 
\mbox{${\varepsilon }_{\rm T}=2$~GHz} 
and tunneling 
\mbox{$\Delta_{\rm T}=25$~MHz},
\mbox{$\tau=10$~ns}, and 
\mbox{$\xi =0.4$~GHz} 
the conditions above are met. 
That would be true even if we change the coupling strength to \mbox{$\xi =0.6$~GHz}. 

\noindent In order to realize an $X$ operation on the target qubit, the effective bias is set to zero on the desired subspaces (rows 1, 2, 4, 7, 8, 11, 13, and 14 in Table \ref{Table Effective Bias}). This results in $E=\varepsilon_{\rm T}\ \pm \ 2\xi =0$ \, leading to
\bea
\label{bias_magnitude} 
{\varepsilon }_{{\rm T}_{1}}=2\xi, \quad                
{\varepsilon }_{{\rm T}_{2}}=-2\xi .                    
\ea
Note that this gate operation also results in a phase factor $3\pi/2$  on the target qubit which can be tracked in the course of computation. As we showed the multi-qubit gate that detects the parity of four can be performed in a sequence of two controlled-unitary operations ($2\tau$) which is faster than detecting the parity of two out of four which takes a sequence of three controlled-unitary operations ($3\tau$).

\section{ Surface Code error syndrome detection based on multi-qubit gates}

\begin{figure}[htp]
    \centering
    \includegraphics[width=2.82in, height=2.82in, keepaspectratio=false]{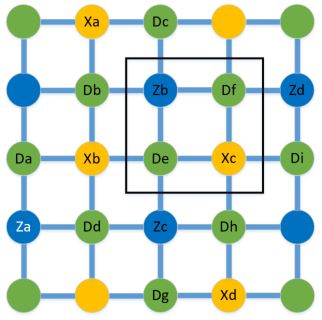}
    \caption{A 2D array of  qubits with nearest-neighbor couplings forming a Surface-17 planar code logical qubit. Here 17 physical qubits (labeled) are required to form a logical qubit, 9 of which are data qubits and 8 of them are measurement ones. The box shows two data qubits De and Df in green, one measure-$Z$ qubit Zb in blue and one measure-$X$ qubit Xc in orange.}
    \label{Fig. 5}
\end{figure}

\indent 
 In Surface Code scheme \cite{fowler_mariantoni_martinis_cleland_2012}, a 2D array of physical qubits is constructed with interleaving data qubits and measurement qubits called measure-$Z$ and measure-$X$ ancillary qubits, and a methodology is presented to protect the architecture from both bit-flip and phase-flip errors at the same time. The measure-$X$ and measure-$Z$ qubits detect phase-flip and bit-flip parities, respectively. As shown in Fig.~\ref{Fig. 5}, each data qubit in Surface Code is surrounded with 4 measurement qubits while each measurement qubit is surrounded with 4 data qubits. At the start, all measurement qubits are initialized to zero. At each error correction cycle, we perform measurements only on ancillary measurement qubits which stabilize the data qubits. Note that the states of data qubits are not perturbed by the measurement. A software maps the detected error syndromes (bit-flip, phase-flip, measurement error) to a graph model which keeps track of errors and fixes the errors \cite{kelly_2015, fowler_mariantoni_martinis_cleland_2012}.

\indent 
For instance, in Fig.~\ref{Fig. 5}, the qubit Zb forces the data qubits Df, De, Dc, and Db to an eigenstate of
${\hat{Z}}_{\rm Df}\, {\hat{Z}}_{\rm De}\,{\hat{Z}}_{\rm Dc}\,{\hat{Z}}_{\rm Db}$ operator, 
while the qubit Xc forces the data qubits Di, Df, Dh, and De to an eigenstate of 
${\hat{X}}_{\rm Di}{\hat{X}}_{\rm Df}{\hat{X}}_{\rm Dh}{\hat{X}}_{\rm De}$ operator. 
Note the chosen stabilizer operators 
${\hat{X}}_{\rm Di}\, {\hat{X}}_{\rm Df}\, {\hat{X}}_{\rm Dh}\, {\hat{X}}_{\rm De}$ and
${\hat{Z}}_{\rm Df}\, {\hat{Z}}_{\rm De}\, {\hat{Z}}_{\rm Dc}\, {\hat{Z}}_{\rm Db}$ must commute with one another to force the projective measurement outcome of the system into a unique eigenstate of all the stabilizers. 
Moreover, the order of applying $\hat{X}$ and $\hat{Z}$ operators on data qubits is important. The order must be chosen to ensure that we are not measuring the result of $\hat{X}$ and $\hat{Z}$ operators of any data qubit simultaneously. Failure to keep the commutation relationship of neighbor stabilizers results in random measurements \cite{fowler_mariantoni_martinis_cleland_2012}. In our example, the order of $\hat{X}$ and $\hat{Z}$ in 
${\hat{X}}_{\rm Di}{\hat{X}}_{\rm Df}{\hat{X}}_{\rm Dh}{\hat{X}}_{\rm De}$ and ${\hat{Z}}_{\rm Df}{\hat{Z}}_{\rm De}{\hat{Z}}_{\rm Dc}{\hat{Z}}_{\rm Db}$ 
guarantees that the two stabilizers are commuting as well as the shared data qubits Df and De between the two stabilizer types 
${\hat{X}}_{\rm Di}{\hat{X}}_{\rm Df}{\hat{X}}_{\rm Dh}{\hat{X}}_{\rm De}$ and  ${\hat{Z}}_{\rm Df}{\hat{Z}}_{\rm De}{\hat{Z}}_{\rm Dc}{\hat{Z}}_{\rm Db}$ are interacting with one ancilla qubit of a type (Xc or Zb) at a time. This ensures the robustness of Surface Code to ancilla errors \cite{tomita_svore_2014}.

\indent The quantum circuits to detect bit-flip error or phase-flip error during one cycle of Surface Code are based on applying CNOT gates. In some systems, one could perform a CNOT gate on any pair of neighboring qubits while the unwanted couplings to the other neighboring qubits are shut off or sufficiently detuned such that their interaction with the target qubit can be neglected. 
However, in systems with always-on interactions, the coupling values can not be tuned or shut off during the error syndrome detection and performing CNOT gates are costly. Here we consider designing new multi-qubit gates to facilitate error syndrome detection in such systems. In this section, we discuss different scenarios to realize Surface Code error correction for systems with always-on interactions using the introduced multi-qubit gates.

\begin{figure}[htp]
    \centering
    \includegraphics[width=2.58in, height=2.58in, keepaspectratio=false]{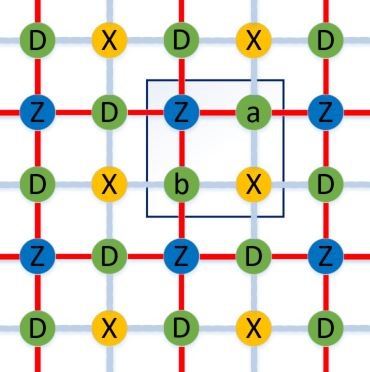}
    \caption{A 2D array of Surface Code where each pair of data qubits share a measure-$X$ qubit and a measure-$Z$ qubit as shown in the square box. All measure-$Z$ qubits are coupled to data qubits with the same coupling strength shown in red, while all measure-$X$ qubits are coupled to data qubits with the coupling strength shown in light grey.}
    \label{Fig. 6}
\end{figure}

\begin{figure*}
\begin{subfigure}{0.31\textwidth}
\includegraphics[width=\linewidth]{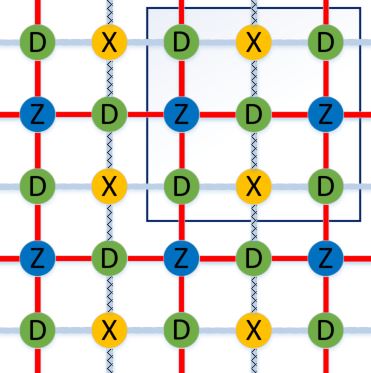}
\subcaption{}
\end{subfigure}
\hspace*{\fill} 
\begin{subfigure}{0.31\textwidth}
\includegraphics[width=\linewidth]{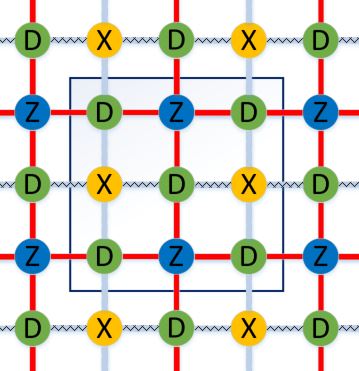}
\subcaption{}
\end{subfigure}
\hspace*{\fill} 
\begin{subfigure}{0.31\textwidth}
\includegraphics[width=\linewidth]{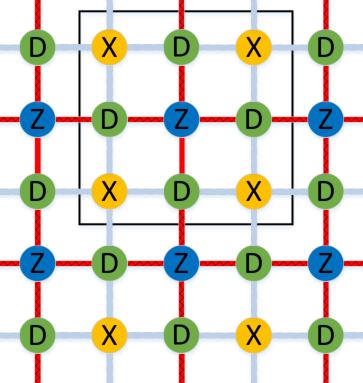}
\subcaption{}
\end{subfigure}
\noindent \caption{\label{SurfaceCode_applying_gates}A two-dimensional array of Surface Code. ~(a) Applying multi-qubit X operators on all vertical columns in 2D array of qubits shown in dotted lines.  ~(b) Applying multi-qubit X operator on all horizontal rows in 2D array of qubits shown in dotted lines. ~(c) Applying multi-qubit Z operators on 2D array of qubits shown in dotted red lines.}
\end{figure*}

\indent Consider a large fabric of Surface Code with the proposed architecture shown in Fig.~\ref{Fig. 6}. Here, all measure-$Z$ qubits are coupled to the surrounded data qubits using the same coupling strength ${\xi }_{\rm A}=0.4$~GHz and all measure-$X$ qubits are coupled to the surrounded data qubits by the same coupling strength of ${\xi }_{\rm B}=0.6$~GHz. In this architecture, we can use our multi-qubit gates plus Hadamard gates to realize one cycle of error syndrome detection. The order of applying multi-qubit gates is given below -- note that Hadamard gates are applied on all measure-$X$ qubits at the beginning and end of each cycle:
\begin{enumerate}[label=\Alph*.]
\item  Apply five-qubit parity gates where data qubits are the target qubits, and top and bottom X stabilizers are the active control qubits, while the left and right Z stabilizers are the dummy qubits (see Fig.~\ref{SurfaceCode_applying_gates}~(a)). 

\item  Apply five-qubit parity gates where data qubits are the target qubits, and top and bottom Z stabilizers are the active control qubits, while the left and right X stabilizers are the dummy qubits (see Fig.~\ref{SurfaceCode_applying_gates}~(b)). 

\item  Apply five-qubit parity gates where all Z stabilizers are the target qubits and all data qubits are the active control qubits (see Fig.~\ref{SurfaceCode_applying_gates}~(c)). 
\end{enumerate}

\noindent

\noindent 

\indent Note that the order of applying these multi-qubit operators is important. 
Next we use the Heisenberg representation \cite{fowler_mariantoni_martinis_cleland_2012} and work on the stabilizer formalism to confirm the correct order by evaluating two different choices. For simplicity, we consider a small subspace of two data qubits a and b, and two measurement qubits Z and X as shown in Fig.~\ref{Fig. 6}. Here, we analyze the effect of our multi-qubit gates acting on the small subspace of interest step by step. 

\indent 
Let us start with the order of A, C,  B from  above; first a multi-qubit vertical $\hat{X}$ operator, second a $\hat{Z}$ operator, third a horizontal $\hat{X}$ operator. Consider the box of four qubits as depicted in Fig.~\ref{Fig. 6}. Initially, the measure-$X$ and measure-$Z$ qubits are initialized to $\left.|+\right\rangle $ and $\left.|0\right\rangle $, respectively, and the system is in a simultaneous eigenstate of the two operator products
$
{\hat{X}}_X\,{\hat{I}}_{\rm a}\,{\hat{I}}_{\rm b}\, {\hat{I}}_Z$ 
and
${\hat{I}}_X\,{\hat{I}}_{\rm a}\, {\hat{I}}_{\rm b}\, {\hat{Z}}_Z $.
There is a tensor product between each pair of single-qubit operators but  is removed for simplicity.
Then applying the vertical $\hat{X}$ operator results in
${\hat{X}}_X\, {\hat{X}}_{\rm a}\,{\hat{I}}_{\rm b}\,{\hat{I}}_Z$
and
${\hat{I}}_X\,{\hat{I}}_{\rm a}\,{\hat{I}}_{\rm b}\, {\hat{Z}}_Z $.
Applying the {$\hat{Z}$} operator leads to
${\hat{X}}_X\,{\hat{X}}_{\rm a}\,{\hat{I}}_{\rm b}{\hat{X}}_Z$
and
${\hat{I}}_X\,{\hat{Z}}_{\rm a}\,{\hat{Z}}_{\rm b}\,{\hat{Z}}_Z$.
And finally applying the horizontal \textit{$\hat{X}$} operator results in
\bea 
\label{Eq. 39}
{\hat{X}}_X\,{\hat{X}}_{\rm a}\,{\hat{X}}_{\rm b}\,{\hat{X}}_Z
\qquad 
{\rm and}
\qquad
{\hat{Z}}_X\,{\hat{Z}}_{\rm a}\,{\hat{Z}}_{\rm b}\,{\hat{Z}}_Z \, .
\ea
 The order chosen above will not work since the resulted stabilizers in Eq.~\ref{Eq. 39} do not commute and the single measurements of $\hat{X}$ and $\hat{Z}$  operators give us random results. 

\indent Next we consider the order of A, B,  C from the above; first a multi-qubit vertical $\hat{X}$  operator, second a horizontal $\hat{X}$ operator, and third a $\hat{Z}$ operator. Initially, we have 
${\hat{X}}_X\,{\hat{I}}_{\rm a}\,{\hat{I}}_{\rm b}\,{\hat{I}}_Z $
and
${\hat{I}}_X\,{\hat{I}}_{\rm a}\,{\hat{I}}_{\rm b}\,{\hat{Z}}_Z $.     
Applying the vertical X operator results in
${\hat{X}}_X\,{\hat{X}}_{\rm a}\, {\hat{I}}_{\rm b}\, {\hat{I}}_Z$
and
${\hat{I}}_X\,{\hat{I}}_{\rm a}\,{\hat{I}}_{\rm b}\, {\hat{Z}}_Z$.
Then applying the horizontal X operator leads to
${\hat{X}}_X\,{\hat{X}}_{\rm a}\,{\hat{X}}_{\rm b}\,{\hat{I}}_Z$
and
${\hat{I}}_X\,{\hat{I}}_{\rm a}\,{\hat{I}}_{\rm b}\,{\hat{Z}}_Z$.                   
Finally,  applying the Z operator results in
\bea 
\label{Eq. 43}
{\hat{X}}_X\,{\hat{X}}_{\rm a}\,{\hat{X}}_{\rm b}\,{\hat{I}}_Z
\qquad
{\rm and}
\qquad
{\hat{I}}_X\,{\hat{Z}}_{\rm a}\,{\hat{Z}}_{\rm b}\, {\hat{Z}}_Z  \, .
\ea
This order of applying multi-qubit gates guarantees that each two data qubits share a pair of $\hat{X}$ and $\hat{Z}$ stabilizers and the measurements after each cycle are valid. 

\noindent In all three steps above, the states of all qubits that are not used in the multi-qubit gate operations are frozen. The error correction cycle is performed by applying 3 sequences of multi-qubit parity gates as ordered in \mbox{Fig.~\ref{SurfaceCode_applying_gates}~(a), ~(b), and ~(c).} We can add/remove an arbitrary number of multi-qubit parity gates to scale up or down these gates in a larger 2D array of qubits when realizing a large-scale Surface Code memory. 

\begin{figure*}
\begin{subfigure}{0.31\textwidth}
\includegraphics[width=\linewidth]{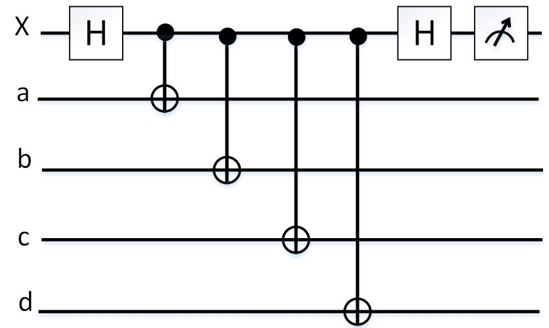}
\subcaption{}
\end{subfigure}
\hspace*{\fill} 
\begin{subfigure}{0.31\textwidth}
\includegraphics[width=\linewidth]{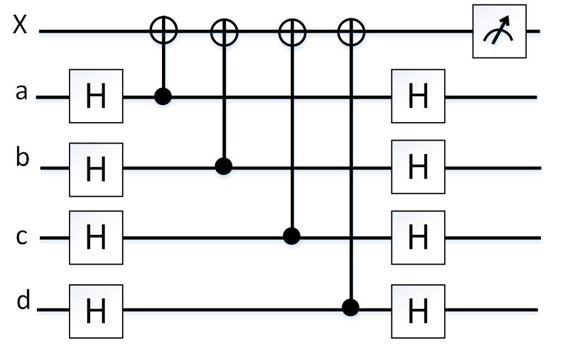}
\subcaption{}
\end{subfigure}
\hspace*{\fill} 
\begin{subfigure}{0.31\textwidth}
\includegraphics[width=\linewidth]{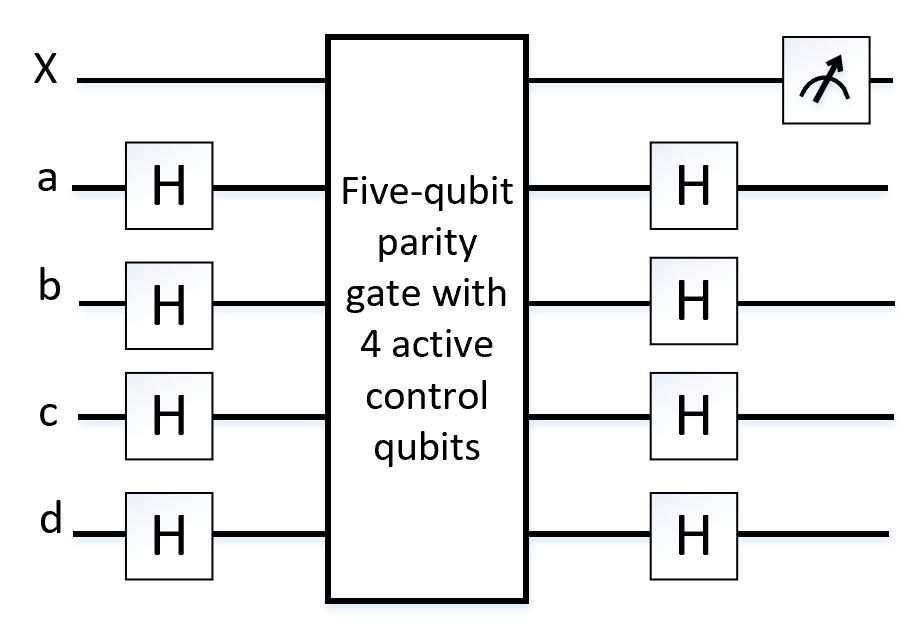}
\subcaption{}
\end{subfigure}
\caption{\label{MQ_stabilizerCircuit}The logically equivalent phase error syndrome detection circuits (a) The conventional phase-error syndrome detection circuit (b) The equivalent phase error syndrome detection circuit where the direction of CNOT gates are reversed and some consecutive pair of Hadamard gates are canceled out (c) The equivalent phase error syndrome detection circuit utilizing the five-qubit parity gate with four active control qubits (a, b, c, d) and X as the target qubit} 
\end{figure*}

\indent The proposed protocol can be further improved by considering a new phase-error-detection circuit utilizing a five-qubit parity detection gate with four active control qubits.
The conventional phase-error-detection circuit is shown in Fig.~\ref{MQ_stabilizerCircuit}~(a). One can reach the same functionality by reversing the direction of each CNOT gate and sandwiching it between Hadamard gates on both control and target qubits. Canceling out the consecutive pair of Hadamard gates on measure-$X$ qubit, the equivalent circuit is shown in Fig.~\ref{MQ_stabilizerCircuit}~(b). Moreover, the functionality of the four CNOT gates shown in Fig.~\ref{MQ_stabilizerCircuit}~(b) can be achieved by a five-qubit parity gate with four active control qubits. As depicted in Fig.~\ref{MQ_stabilizerCircuit}~(c), to realize a phase-flip detection circuit, first, we apply  Hadamard gates on data qubits surrounding the measure-$X$ qubit, then we apply a five-qubit parity gate with four data qubits acting as active control qubits and the measure-$X$ qubit acting as the target qubit. Finally, we apply Hadamard gates on the four data qubits. 
\noindent

\indent 
Utilizing the five-qubit parity gates with four active control qubits plus single-qubit Hadamard gates, one can realize Surface Code error detection cycles in a sequence of only two multi-qubit gates plus single-qubit gates. Here the order of applying multi-qubit gates is not important and Hadamard gates are applied on all data qubits at the beginning and end of each cycle.

\begin{enumerate}[label=\Alph*.]
\item  Apply five-qubit parity gate where the X stabilizers are the target qubits and the surrounded data qubits are the active control qubits.

\item  Apply five-qubit parity gates where the Z stabilizers are the target qubits and the surrounded data qubits are the active control qubits. 
\end{enumerate}

 \section{Discussion}

\noindent

\begin{figure*}
\begin{subfigure}{0.31\textwidth}
\includegraphics[width=\linewidth]{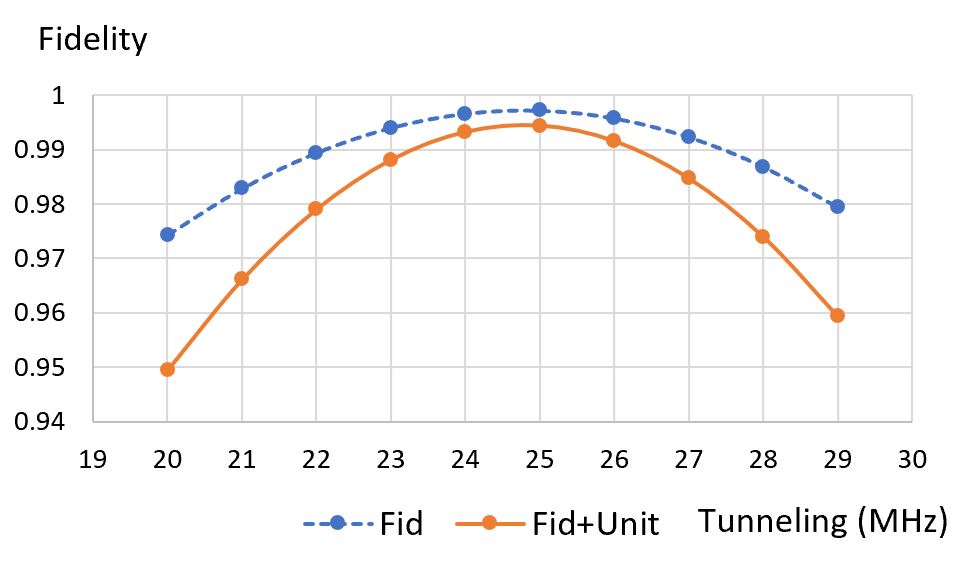}
\subcaption{}
\end{subfigure}
\hspace*{\fill} 
\begin{subfigure}{0.31\textwidth}
\includegraphics[width=\linewidth]{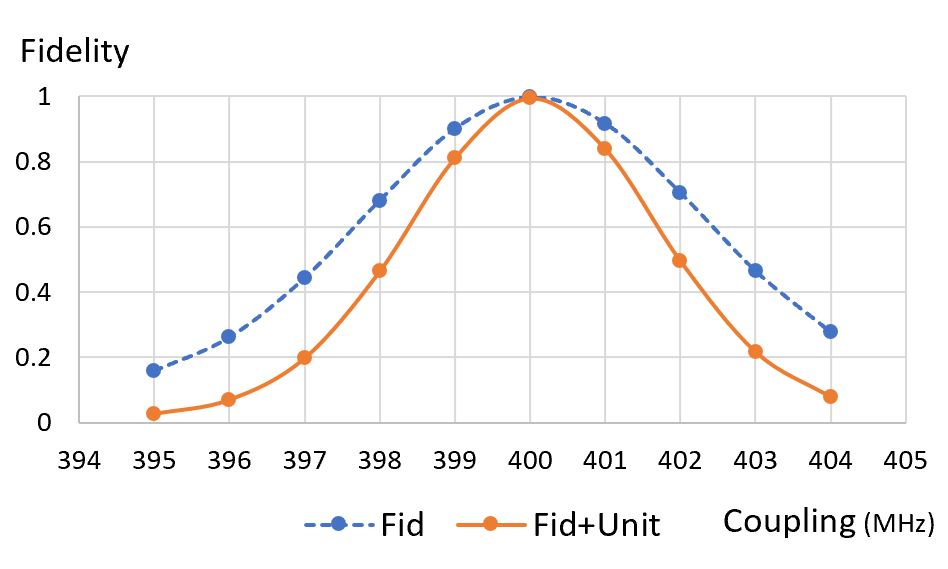}
\subcaption{}
\end{subfigure}
\hspace*{\fill} 
\begin{subfigure}{0.31\textwidth}
\includegraphics[width=\linewidth]{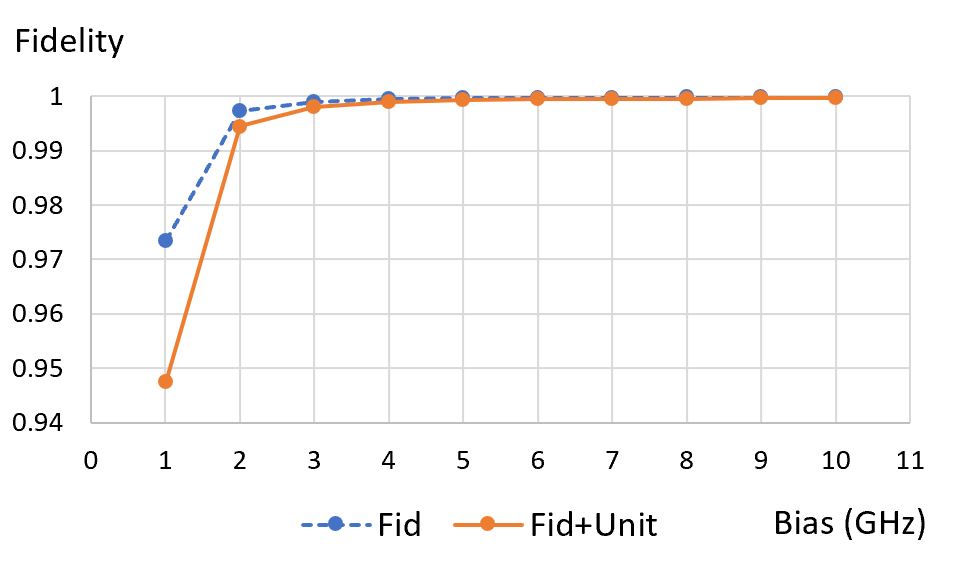}
\subcaption{}
\end{subfigure}
\caption{\label{Gate_Sensitivity}The effect of changing the system parameters on the fidelity of the parity gate with four active control qubits. Initial parameters are set as $\Delta_{\rm T}=25$~MHz, $\tau= 10$~ns, $\xi= 0.4$~GHz, and bias on control qubits as 2~GHz, then one parameter (tunneling, coupling or bias) is changed while all others are constant. Here we considered two different fidelity formulas as discussed in the main text. Fid represents the fidelity based on Eq.~\ref{Fidelity}, and Fid+Unit represents the fidelity  based on Eq.~\ref{Fidelity_Unitary} where the unitary condition of the gate is also evaluated.  (a) The effect of changing the tunneling (b) The effect of equally changing the coupling strengths (c) The effect of changing the bias on control qubits.} 
\end{figure*}

We use our derived gate parameters in a MATLAB simulator that performs time evolution of a nine-qubit system as shown in Fig.~\ref{Fig. 2}. The simulator solves the Schr\"{o}dinger equation based on trotterization \cite{hatano_suzuki_2005} method with 0.1 ns trotter steps . We consider qubits E, F, G, H in the simulation to show that their states remain unchanged during the five-qubit gate operations. We use the following equations for calculating the gate fidelity:
\bea
\label{Fidelity} 
{\rm Fid}= \frac{\abs{\, {\rm Tr}\left(U_{\rm ideal }^\dagger\, U\right)\,}}{d} \ {\rm ,} 
\ea

\bea
\label{Fidelity_Unitary} 
{\rm Fid+Unit} = \frac{  {\rm Tr}\left(U^\dagger U\right) + {{\abs{\, {\rm Tr}\left(U_{\rm ideal }^\dagger\, U\right)\,}}^2}}{d\times \left(d+1\right)} 
\ea
where $d=2^9$ is the dimentionality of the computational space,  $U_{\rm ideal}$ is the unitary transformation of the desired ideal gate, and U is the achieved unitary transformation calculated from the time evolution of the system:
\bea 
\label{Unitary_exp} 
U=e^{-i\int^{\tau_{\rm total}}_0{H(t)dt }} \, ,
\ea
with $\tau_{\rm total}$ being the overall duration of the gate operation and $H(t)$ being the Hamiltonian of the system at time $t$. The fidelity equation in Eq.~\ref{Fidelity_Unitary} accounts for checking the unitary condition of the quantum operation \cite{pedersen_moller_molmer_2007} and reports slightly lower gate fidelity as it is depicted in Fig.~\ref{Gate_Sensitivity}. 

\indent 
Note that in an experimental setup, one can realize the presented gates by choosing a different set of parameters which match with their physical system. One may  choose different integers or multiply each parameter by a scaling factor such that the conditions explained in sections \ref{ParityGateTwoControlQ} and \ref{ParityGateFourControlQ} remain satisfied \cite{kumar_skinner_Behrman_steck_Han_2005, kumar_daraeizadeh_2015}. For example, another set of parameters satisfying Eq.~\ref{cos-sin=0} would be $\Delta_{\rm T}=25$~MHz, $n= 1$, and $\tau= 50$~ns. Or the same gate fidelity can be achieved by keeping $n=0$, but changing $\tau$ to 20~ns and reducing the tunneling parameter to $\Delta_{\rm T}=12$~MHz instead. An example of implementing controlled-unitary gates deriven by bias pulse scheme on an rf SQUIDs physical system has been presented in Ref.~\cite{kumar_skinner_Behrman_steck_Han_2005}. The parameters such as tunneling, coupling, bias pulse magnitude and duration chosen in this paper are in the same range as discussed in  Ref.~\cite{kumar_skinner_Behrman_steck_Han_2005}, where it is shown how these parameters can be adjusted to realize the controlled-unitary gates on the hardware. 

\indent Figure \ref{Gate_Sensitivity} shows the sensitivity of the parity gate with four active control qubits on the different parameters. As depicted in Fig.~\ref{Gate_Sensitivity}~(a), a mismatch of up to 2~MHz in the tunneling value results in the fidelity drop of \mbox{$<\,1$\%}, however this can be compensated by adjusting the bias pulse width $\tau$ on the target qubit. In Fig.~\ref{Gate_Sensitivity}~(b) we swept away the coupling value of all four control qubits from the designed value $\xi=0.4$~GHz and plotted the fidelity change. As it can be seen, if we use the same bias pulse magnitudes from Eq.~\ref{bias_magnitude}, ${\varepsilon }_{{\rm T}_{1}}=0.8,\ {\varepsilon }_{{\rm T}_{2}}=-0.8$, the gate fidelity drops significantly. However, if we change the magnitudes of the bias pulse according to the new $\xi$ values, we can achieve a high fidelity gate again. As we discussed, any error from the parameter mismatch in tunneling and couplings can be respectively recovered by adjusting the bias pulse duration and magnitude. Therefore, the control circuitry is greatly reduced since by only adjusting one control parameter (bias pulse), one can achieve a high fidelity gate. 

\indent Our simulation shows that increasing the chosen bias value on control qubits would result in better gate fidelity. In Fig.~\ref{Gate_Sensitivity}~(c) the resulted fidelity vs bias values varying from 1~GHz to 10~GHz is plotted. For instance, with bias on control qubits as 2~GHz, the fidelity of the parity gate with four active control qubits (with \mbox{$\Delta_{\rm T}=25$~MHz}, $\xi= 0.4$~GHz, and $\tau= 10$~ns), was 0.9972 and 0.9944 based on Eq.~\ref{Fidelity} and Eq.~\ref{Fidelity_Unitary}, respectively. However, changing the bias on control qubits to 3 GHz resulted in gate fidelity of 0.999 and 0.998, respectively.

\indent In the physical realizations, the bias pulses are not ideal and have some rise/fall times depending on the control electronics. The effect of the rise/fall times can be compensated by slightly changing the gate duration times \cite{kumar_skinner_Behrman_steck_Han_2005}. Ideally one can use the analytical methods to design the ideal bias pulses and then use optimization methods to optimize the rise/fall times and bias pulse shapes based on their physical system to achieve the highest fidelity. 

\indent In this work, we considered the Hamiltonian with Ising interactions, however, the proposed gates can be realized for Hamiltonians with XX and YY interactions by simply interchanging the tunneling and bias values while coupling values and other parameters remain unchanged \cite{kumar_daraeizadeh_2015, kumar_skinner_2007}. Furthermore, here we considered an arbitrary size 2D array of qubits to represent the application of multi-qubit parity gates in Surface Code schemes. However, it is often required to perform a reduced $\hat{X}$ or $\hat{Z}$ stabilizers on the borders of a logical qubit. To realize a two-terminal stabilizer, one can use the five-qubit parity gate with two active controls. Also realizing a three-terminal stabilizer is possible using a five-qubit parity gate with three active control qubits. Designing a five-qubit parity gate with three active control qubits using the methods discussed here is straightforward. Note that different coupling strengths are engineered depending on the number of the active control qubits in a multi-qubit gate and this effects on the architectural design decisions of the Surface Code array in systems with always-on interactions.

\noindent

\section{ Conclusion}

\indent We designed new five-qubit parity gates with the fidelity of \mbox{$>\,99.9$\%} for nearest-neighbor architectures with always-on Ising interactions. There are many applications for these new gates, such as performing quantum state transfer in blocks of two-dimensional (2D) array of qubits. In this paper, we utilized these gates in error-syndrome-detection circuits. We designed a new quantum memory architecture for systems with always-on interactions, and presented a Surface Code protocol based on multi-qubit gates. The five-qubit parity gates can simultaneously be applied on many qubits in the array of Surface Code by adjusting only one control parameter (bias on the target qubits). Here, the Surface Code cycles can be achieved by applying two sequences of five-qubit parity gates across the entire qubit array, with the duration of each sequence being $2\tau$ plus the timing required for single-qubit gates and measurements. The conventional Surface Code schemes based on two-qubit gates use the same timing for single-qubit gates and measurements, however, they need at least three sequences of CNOT gates across the qubit array. In the 2D qubit systems with always-on interactions, each CNOT gate takes $8\tau$ which adds up to $24\tau$ for a full Surface Code cycle. The advantages of using our proposed Surface Code memory architecture can be summarized in four main points:

\begin{enumerate}
\item  It extensively simplifies the control circuitry. 
\item  It achieves a much faster error-correction cycle compared to the error syndrome detection circuits based on two-qubit gates. 
\item  It is expandable to large-scale Surface Code architectures with a fixed circuit depth for any size of a 2D array of qubits. 
\item  It removes the possibility of developing relative phases (dephasing) during idle times since there are no idle qubits in this scheme.
\end{enumerate}

\noindent 

\noindent 

\bibliographystyle{unsrt}


\noindent

\end{document}